%% file: conference_101719.tex
\documentclass[10pt,conference]{IEEEtran} 
\IEEEoverridecommandlockouts
\usepackage{etoolbox}
\AtBeginEnvironment{tabular}{\centering}

\usepackage{cite}
\usepackage{amsmath,amssymb,amsfonts}
\usepackage{algorithmic}
\usepackage{graphicx}
\usepackage[affil-it]{authblk} 
\usepackage[breaklinks=true]{hyperref}
\usepackage{tabularx}
\usepackage{xurl}
\usepackage{longtable}
\usepackage{hyperref}
\usepackage{textcomp}
\usepackage{threeparttable}
\usepackage{float}
\usepackage{bbding}
\usepackage{pifont}
\usepackage{ragged2e}
\usepackage{amssymb}
\usepackage[utf8]{inputenc}
\usepackage{multirow}
\usepackage[utf8]{inputenc}
\usepackage{booktabs} 
\usepackage{ltablex}
\usepackage{makecell} 
\keepXColumns 
\usepackage{longtable} 
\usepackage{color}
\usepackage{tabularray}
\usepackage{authblk}
\usepackage[utf8]{inputenc}
\usepackage{array}
\usepackage[margin=1in]{geometry}
\usepackage{listings}
\usepackage{xcolor}
\usepackage{xspace}
\usepackage{url}
\usepackage[most]{tcolorbox}
\usepackage{hyperref} 
\usepackage[T1]{fontenc} 
\usepackage{geometry} 
\def\BibTeX{{\rm B\kern-.05em{\sc i\kern-.025em b}\kern-.08em
    T\kern-.1667em\lower.7ex\hbox{E}\kern-.125emX}}
\begin{document}
\newcommand{\todo}[1]{\textbf{\textcolor{red}{ToDO: #1}}}
\newcommand{\wei}[1]{\textbf{\textcolor{orange}{Wei: #1}}}
\newcommand{\recheck}[1]{\textcolor{green}{ #1}}
\newcommand{\fix}[1]{\textcolor{blue}{ #1}}
\newcommand{\revision}[1]{\textcolor{blue}{ #1}}

\newcommand{\answer}[2]{
  \begin{tcolorbox}[enhanced, left=3mm,right=3mm,
    colback=gray!10, colframe=gray!80, boxrule=0pt,
    borderline west={4pt}{0pt}{gray!90},
    ]
    \textbf{Answer for RQ#1:}
    #2
    \end{tcolorbox}
}

\def\figref#1{Figure~\ref{fig:#1}}
\def\figlabel#1{\label{fig:#1}\label{p:#1}}
\def\tabref#1{Table~\ref{tab:#1}}
\def\tabsref#1{Tables~\ref{tab:#1}}
\def\tablabel#1{\label{tab:#1}\label{p:#1}}
\def\Secref#1{\S\ref{sec:#1}}
\def\secref#1{\S\ref{sec:#1}}
\def\seclabel#1{\label{sec:#1}}
\def\qref#1{Eq.~\ref{eqn:#1}}
\def\eqrefn#1{\ref{eqn:#1}}
\def\eqsref#1#2{Eqs.~\ref{eqn:#1}/\ref{eqn:#2}}
\def\eqlabel#1{\label{eqn:#1}}
\def\subsp#1{P_{\mbox{{\scriptsize\rm #1}}}}
\def\appref#1{Appendix~\ref{sec:#1}}

\title{SoK: Comprehensive Analysis of Rug Pull Causes, Datasets, and Detection Tools in DeFi}



\author[1]{Dianxiang Sun}
\author[1]{Wei Ma}
\author[2]{Liming Nie}
\author[1]{Yang Liu}

\affil[1]{Nanyang Technological University, Singapore}
\affil[2]{Shenzhen Technology University, China}

\maketitle

\begin{abstract}
Rug pulls pose a grave threat to the cryptocurrency ecosystem, leading to substantial financial loss and undermining trust in decentralized finance (DeFi) projects.
With the emergence of new rug pull patterns, research on rug pull is out of state. To fill this gap, we first conducted an extensive analysis of the literature review, encompassing both scholarly and industry sources. By examining existing academic articles and industrial discussions on rug pull projects, we present a taxonomy inclusive of 34 root causes, introducing six new categories inspired by industry sources: \textit{burn}, \textit{hidden owner}, \textit{ownership transfer}, \textit{unverified contract}, \textit{external call}, and \textit{fake LP lock}. Based on the developed taxonomy, we evaluated current rug pull datasets and explored the effectiveness and limitations of existing detection mechanisms. Our evaluation indicates that the existing datasets, which document 2,448 instances, address only 7 of the 34 root causes, amounting to a mere 20\% coverage. It indicates that existing open-source datasets need to be improved to study rug pulls. In response, we have constructed a more comprehensive dataset containing 2,360 instances, expanding the coverage to 54\%  with the best effort. In addition, the examination of 14 detection tools showed that they can identify 25 of the 34 root causes, achieving a coverage of 73.5\%. There are nine root causes~(\textit{Fake LP Lock}, \textit{Hidden Fee}, and \textit{Destroy Token}, \textit{Fake Money Transfer}, \textit{Ownership Transfer}, \textit{Liquidity Pool Block}, \textit{Freeze Account}, \textit{Wash-Trading}, \textit{Hedge}) that the existing tools cannot cover. Our work indicates that there is a significant gap between current research and detection tools, and the actual situation of rug pulls.
\end{abstract}


\input{introduction}
\input{background_and_motivation}
\input{methodology}

\input{results_and_insights}

\input{discussion}

\input{related_work}
\input{conclusion}

\bibliographystyle{IEEEtran}
\bibliography{reference}
\end{document}

%% file: introduction.tex
\section{Introduction}
The emergence of decentralized finance (DeFi) has been transformative for the financial sector, introducing innovative solutions and creating new opportunities. However, this rapidly evolving ecosystem is not without its perils, as it has become fertile ground for fraudulent schemes, among which rug pulls pose a substantial threat. Rug pulls occur when developers launch a new cryptocurrency token, draw in investors, and subsequently abandon the project, absconding with the funds and rendering the tokens valueless \cite{Do_Not_Rug_on_Me, An_analysis_of_crypto_scams_during_the_Covid-19_pandemic, Rug_pull_malicious_token_detection_on_blockchain_using_supervised_learning_with_feature_engineering}. The frequency of these deceptions has resulted in significant financial losses, eroding the trust in DeFi projects and impeding the broader acceptance of decentralized technologies.

Scholarly efforts to study rug pulls have predominantly centred on two dimensions: empirical analysis and the development of detection strategies. Empirical investigations have examined the patterns and affection of rug pulls on various blockchain platforms\cite{Scam_Alert,From_Programming_Bugs_to_Multimillion_Dollar_Scams,kaur2023risk}, providing critical insights for rug pulls. Conversely, the current detection methodologies employ machine learning, fuzzing, or hybrid analytical frameworks designed to pre-emptively identify and thwart such frauds\cite{TokenAuditor}\cite{A_Deep_Dive_into_NFT_Rug_Pulls}\cite{Rug_pull_malicious_token_detection_on_blockchain_using_supervised_learning_with_feature_engineering}.

However, the sphere of research in this domain is not without its shortcomings. A primary limitation is the lack of uniformity and completeness in the taxonomies available for classifying rug pull schemes, exacerbating the challenge of devising a consolidated and all-encompassing taxonomy framework. Furthermore, questions persist regarding the comprehensiveness and representativeness of the datasets employed to support rug pull research, which is essential for the refinement of robust detection methodologies. Furthermore, the efficacy of existing detection tools has not been sufficiently examined, making it challenging to evaluate their success in identifying and counteracting these deceptions. 

To address these gaps, we have developed the Rug Pull Analysis Framework (RPAF), a systematic method that includes a thorough literature review, taxonomy creation, dataset critique and reconstruction, and assessment of detection tools. We use seven databases — Google Scholar, IEEE Xplore, Scopus, ACM Digital Library(DL) and arXiv for the academic search; Google Search and Twitter for the grey literature search. We amassed 661 academic papers and the first 50 pages of industrial-report search results until September 30, 2023. From this collection, we rigorously selected 31 academic and 27 industrial reports that are included in our study. We crafted a rug pull taxonomy delineating 34 root causes, then appraised three existing rug pull datasets against this taxonomy. To enhance the data diversity, we expanded the datasets with more rug-pull types. Next, we methodically gathered and assessed 14 rug-pull detection tools using our reconstructed dataset and discovered that none could address all identified root causes. Our research brings to light the nascent understanding of rug pulls, the insufficient breadth and diversity of existing datasets, and the inadequate capability of prevalent detection tools to counteract a range of rug pull strategies. In conclusion, our work contributes threefold:

\begin{itemize}
    \item We develop the most comprehensive taxonomy of rug pulls to date, encompassing 34 root causes categorized into 6 high-level categories and 19 root cause categories, providing a standardized framework for understanding and classifying rug pulls.
    \item We assess the coverage of current rug pull datasets and reconstruct a more comprehensive dataset by integrating existing datasets with 90 real-world rug pull instances, increasing the coverage of the taxonomy from 20\% to 56\%.
    \item We systematically evaluate the capabilities and limitations of 14 current rug pull detection tools, revealing that none of them can cover all root causes in the taxonomy, with the highest coverage being 73.5\%, and demonstrate the necessity for developing a new detection approach.
\end{itemize}

The paper is organized as follows. Section II presents the motivation behind our study. Section III describes our research methodology, including the systematic literature review process, taxonomy construction, dataset evaluation and reconstruction, and tool assessment. Section IV presents the results and insights derived from our analysis, addressing the three research questions. Section V discusses the internal and external threats. Section VI reviews related work in the field of rug pull research. Finally, Section VII concludes the paper. We make our dataset available at [\url{https://github.com/Dianxiang111/Rug-Pull-Dataset}]






%% file: background_and_motivation.tex
\section{ Motivation}

\begin{figure}
    \centering
    \includegraphics[width=0.5\textwidth]{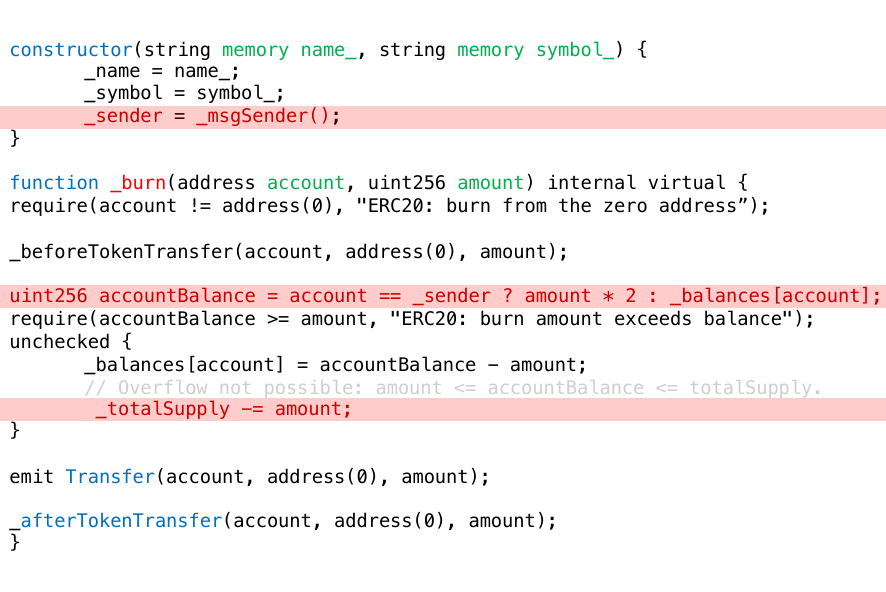}
    \vspace{-2em}
    \caption{Code Backdoor for \$DOGE3.0}
    \label{fig:motivation_example}
\end{figure}

Although several studies have already delineated the taxonomy and patterns of rug pulls, existing research has become increasingly detached from industry practices. Our study intends to address this gap. For example, the rug-pull method employed in the meme-token Dogecoin 3.0 (\$DOGE3.0)~\cite{DOGE3.0} is not encapsulated by any existing taxonomy nor detectable by current open-source or commercial tools. \figref{motivation_example} illustrates the code backdoor implemented for the rug pull. The project has over 300 investors, but the project ultimately proved to be a fraudulent scheme, causing users to lose close to \$70,000. 
To execute the rug pull, firstly, the scammer deployed a token with the proxy mechanism. Later, the scammer hiddenly stored the owner address in the "\_sender" variable as shown by the first red line in \figref{motivation_example}. Secondly, the scammer called the hidden function ``\_burn" to mint tokens as the owner to bypass balance checks and burn tokens without sufficient balance as shown in the second red line in \figref{motivation_example}. Moreover, at the end of the "\_burn" function, the "\_totalSupply" is reduced by the burned token amount as shown in the third red line in \figref{motivation_example}. This increases the total token supply. Eventually, \$DOGE3.0 became a worthless token.


The \$DOGE3.0 rug pull case reveals the limitations of the current taxonomy. The case employs novel techniques such as using proxy contracts to hide the true owner, incorporating hidden minting functions in the contract, bypassing contract restrictions through incorrect account balance checks, and manipulating the total supply to create tokens out of thin air. These methods are not covered by the existing taxonomy. Therefore, to more comprehensively understand and address these evolving rug pull techniques, it is necessary to expand and refine the current taxonomy by including these new categories and characteristics. Only by keeping the taxonomy up-to-date can we better identify risks and protect investors' interests.



%% file: methodology.tex
\section{Research Methodology}

\begin{figure*}[htbp]
\centering
\includegraphics[width=\textwidth]{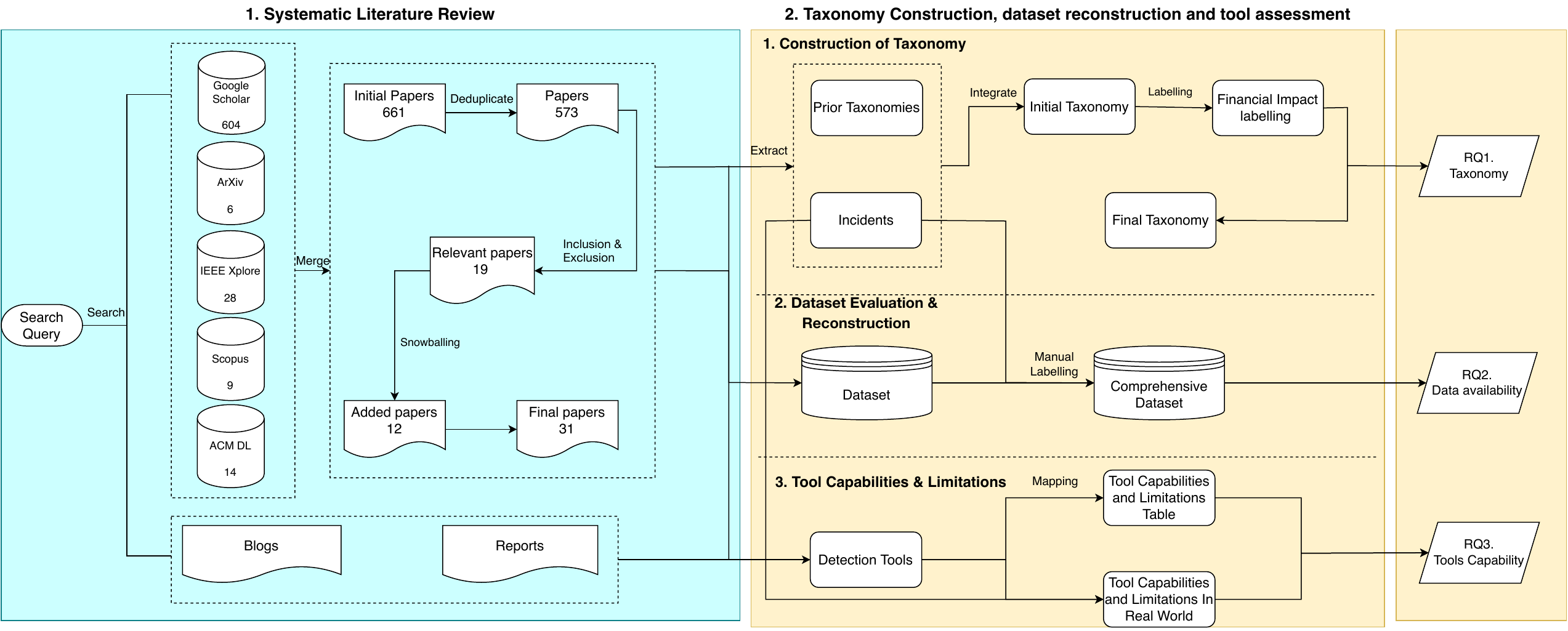}
 \vspace*{-2em}
\caption{RPAF Framework}
\label{fig:overview}
 \vspace*{-2em}
\end{figure*}

In this section, we introduce our research methodology framework. \figref{overview} shows our rug pull analysis framework (RPAF). Our research framework consists of two stages~(systematic literature review and analysis design): the first involves the systematic collection of academic research papers~(scientific literature review) and industry reports on rug pulls~(grey literature review); the second is the analysis process based on the collected articles and reports. We extract the preliminary rug pull taxonomy from the collected materials and then expand it based on the new rug pull cases. Then, we employ the developed taxonomy to evaluate the current rug pull datasets and augment the datasets with new rug-pull patterns. In the end, we evaluate the detection tools based on the re-constructed dataset and our developed taxonomy.



\subsection{Systematic Literature Review}
\subsubsection{Scientific Literature Review}\label{section 3.b}
This section presents a systematic review of academic literature pertaining to research on rug pulls.
We employed a structured approach to carry out this literature review, covering 6 steps of the process\cite{r1,r5} as shown below.

\textit{\textbf{Step 1} - Search Strategy:}  
The search period begins in 2008, as this marks the publication of the original paper describing blockchain technology\cite{nakamoto2008bitcoin}, and concludes on September 30, 2023. 
Our research encompassed only English-language academic journal articles, conference presentations, technical dissertations, as well as postgraduate theses. 
To initiate our investigation, inspired by these works\cite{Do_Not_Rug_on_Me,r3,r4}, we employ the following search queries: 
(``\textit{rugpull}" OR ``\textit{rug pull}" OR ``\textit{exit scam}") AND (``\textit{blockchain}" OR ``\textit{cryptocurrency}" OR ``\textit{decentralized finance}" OR ``\textit{smart contract}") AND (``\textit{detection}" OR ``\textit{prevention}" OR ``\textit{security}" OR ``\textit{vulnerability}" OR ``\textit{attack}" OR ``\textit{backdoor}" OR ``\textit{trap}").

\textit{\textbf{Step 2} -  Literature Search:} A comprehensive literature search was performed utilizing five prominent databases: IEEE Xplore, Scopus, Google Scholar, ACM Digital Library, and arXiv. The search targeted titles, abstracts, and the index terms. 


\textit{\textbf{Step 3} - Merge Candidate Literature:} After conducting our systematic literature search, we documented the quantities of candidate literature retrieved from each database. The distribution was as follows: Google Scholar (604), IEEE Xplore (28), Scopus (9), ACM Digital Library (14), and arXiv (6), totaling 661 initially identified candidate papers.

\textit{\textbf{Step 4} - Deduplication:} 
We conducted a thorough exercise to remove duplicate entries from our initial collection of literature. This careful vetting process resulted in a refined selection of 573 papers. 

\textit{\textbf{Step 5} - Paper Filtering:} The papers collected based on keywords may not necessarily be relevant to our research survey, hence we need further paper filtering. To achieve this goal, we have developed a set of inclusion and exclusion criteria based on a thorough examination of the titles, abstracts, and full texts, to ensure the objectivity of our research results.

To qualify for \textbf{inclusion}, a paper needed to have its title and keywords explicitly citing ``rugpull" OR ``rug pull" OR ``exit scam" or any synonyms which align with our defined keywords. 
Furthermore, the abstract and the full text of the paper must be pertinent to identifying rug pulls or general ideas surrounding them. 
On the other hand, the \textbf{exclusion} criteria stipulated that papers that are inaccessible or unobtainable would be excluded from the analysis. Additionally, studies that discuss blockchain-based fraud but do not provide any substantive discourse on rug pulls were also omitted. The conscientious application of these strict inclusion and exclusion criteria resulted in a significant contraction of the initial array of 573 papers, culminating in a concise, highly relevant corpus of 19 papers. 


\textit{\textbf{Step 6} - Snowballing:} 
In screening the selected literature, we adopted the snowballing method to expand our collection scope and prevent missing important literature. The expansion criteria adhere to the strict inclusion and exclusion standards previously outlined. After careful examination, we added 12 relevant works. These works met our established criteria and were incorporated into our academic collection. In the end, we had 31 papers on rug pull.

\subsubsection{Grey literature review} 
New rug pulls always precede academic research. Thus, we conducted an in-depth study of grey literature, which includes, but is not limited to, technical blog posts, authoritative industry white papers, and comprehensive event reports. To ensure the inclusion of high-quality data sources, we only considered well-known news aggregators and reputable security companies such as Certik, Beosin, DeFi Security, and Peckshield, as well as blogs from Web3 domain security audit experts. Our primary search was conducted through Google and Twitter. 
The search criteria adopted reflect the methods stipulated in Section \ref{section 3.b} of our research framework. This phase was completed when the search no longer produced new insights. In this stage, we ultimately collected 27 pieces of grey literature.


\subsection{\textbf{Taxonomy Construction}}
We employ a three-step process to develop our taxonomy. Initially, we establish a preliminary ``\textit{rug pull}'' taxonomy, drawing upon existing frameworks. Subsequently, we expand this preliminary taxonomy through the analysis of grey literature. Ultimately, to gauge the importance of each category within the developed taxonomy, we quantify their impact in terms of financial losses.

\subsubsection{Building a Preliminary Rug Pull Taxonomy} This initial phase focuses on extracting and analyzing existing taxonomies from academic literature. The process involves manually extracting relevant rug pull types and their taxonomies from 31 academic papers. Using the advanced taxonomy framework proposed by Mazorra et al\cite{Do_Not_Rug_on_Me}. as a reference point, similar rug pull types are integrated based on \textit{textual similarity} (grouping types with identical definitions, references to the same case, or interchangeable terms) and \textit{conceptual similarity} (identifying types that exhibit underlying similarities in nature or modus operandi despite different terminologies). This step results in a preliminary taxonomy, capturing 15 distinct types and 29 fundamental root causes.

\subsubsection{Expanding the Taxonomy} 
The second phase involves integrating insights from grey literature. We reviewed 27 documents of grey literature to identify additional fundamental categories of rug-pull causes. This investigation introduced six new categories derived from the grey literature into the taxonomy, distinguished by asterisks in \tabref{Taxonomy}. This extension aims to incorporate new categories and root causes of rug pull that had not been acknowledged in previous academic research, thus broadening the scope of the taxonomy.

\subsubsection{Quantifying Financial Losses} 
We annotated each rug pull type and associated financial loss amounts in the taxonomy. This involved collecting loss amount reports from case studies in the grey literature and combining them with attacks recorded in the Rekt~\cite{rekt-database} to investigate the monetary losses users experienced in each type of rug pull. Cases with unclear definitions or overlapping categories are classified by detailed discussions and voting among three authors. The quantified total financial impact in the taxonomy reached \$171,713,065, with the token distribution category incurring the most significant economic loss, accounting for 35\% of the total losses.


\subsection{\textbf{Dataset Evaluation and Reconstruction}}
To understand the quality of the rug pull datasets used by the research community, we meticulously collected and used our taxonomy to evaluate these datasets. Additionally, we have gathered a new batch of rug pull data for data augmentation. Augmenting the dataset is crucial for capturing the full spectrum of rug pull incidents, many of which may not be represented in existing academic datasets. 

\subsubsection{Assessment of Existing Datasets} 
To understand the coverage of existing datasets on rug pulls, we investigated 3 existing public datasets from collected papers. There are two main criteria for dataset selection: a) the dataset must explicitly contain contract addresses that have been tagged for the rug pulls type, and b) the dataset needs to be publicly accessible. 
We summarized the critical details of these datasets, including their sources, platforms, size, diversity of rug pull types, and the fundamental causes covered. Then, we classified the rug pull instances from these datasets according to the developed taxonomy. 

\subsubsection{Collection of Real-World Instances} To address gaps in existing datasets, this phase involves gathering real-world instances of rug pulls from grey literature. The selection is guided by two criteria: 1) instances should originate from reputable security companies or blogs from security audit experts in the Web3 domain, and 2) instances should encompass detailed method of operation analysis and corresponding addresses. 90 real-world rug pull instances are collected, each accompanied by a detailed analysis of the fraudulent operations. The main information of the instances, including contract addresses, associated platforms, rug pull types, root causes, loss amounts, and sources of information, is annotated. This step aims to enhance the dataset with detailed and verifiable cases of rug pull fraud.

\subsubsection{Creation of a Comprehensive Rug Pull Dataset} We merge the assessed datasets with the newly collected real-world instances, removing duplicates to form a comprehensive dataset. The three existing datasets collectively encompass 2,508 instances of rug pulls but only cover 7 out of 34 identified rug pull root causes (20\% overall coverage). The reconstructed dataset comprises 13 root cause categories with a total of 2,360 instances after removing the duplicated addresses, increasing the variety of rug pull root causes covered from 7 to 19 types and boosting the overall coverage rate from 20\% to 56\%. This enriched dataset aims to provide a more accurate representation of the diversity and prevalence of rug pulls.


\subsection{\textbf{Tool Capabilities and Limitations}}
We have comprehensively collected detection tools for rug pulls, and evaluate their capabilities and limitations. The evaluation of detection tools is designed to identify strengths and weaknesses in the current technological approaches to identifying rug pull fraud. By methodically comparing these tools against our proposed taxonomy and testing their performance on real-world instances, the study pinpoints gaps in detection capabilities. 

\subsubsection{Tool Identification} During the literature review, we found 58 articles is conducted to identify tools capable of detecting rug pulls in cryptocurrency. Four specific criteria are established for tool selection: 1) relevance to rug pull detection, 2) detailed documentation, 3) full support for our reconstructed datasets, including platforms and input methods, and 4) open accessibility. Applying these criteria, 14 tools are selected from 22 potential candidates for in-depth examination. This step ensures that only tools with potential efficacy in detecting rug pulls are considered for further analysis.

\subsubsection{Assessing Tools' Capabilities} The selected tools are assessed by comparing their detection features against the developed taxonomy of rug pull types and root causes. Each tool's documentation and the indicators it uses to detect rug pulls are thoroughly reviewed. The types of rug pulls each tool can detect are studied and matched against the taxonomy. In instances where a tool's detection capabilities differ from the taxonomy's descriptions, extensive discussions and voting among the four authors are conducted to reach a consensus. This assessment helps understand how comprehensively current tools can identify various forms of rug pull fraud. The 14 tools collectively identify 25 out of 34 Rug Pull root causes, achieving a coverage rate of 73.5\%. However, there are still 9 rug pull root causes (accounting for 36\% of the total) that all tools fail to detect.

\subsubsection{Empirical Testing of Tools} The final phase involves testing the selected tools on 30 verified instances drawn from the augmented dataset to gauge their precision under practical conditions. 15 real-world instances labeled with the highest financial losses within their respective root cause categories and 15 instances involving combination operations leading to significant financial losses are chosen. The detection outcomes of the tools in these instances are evaluated to explore their practical capabilities in real-world conditions.

\subsection{Research Questions(RQs)}
In this study, we have three research questions as follows.
\subsubsection{\textbf{RQ1}} \textit{What is the taxonomy of root cause of rug pulls?} A detailed and universally recognised taxonomy is essential for the effective comprehension, detection, and prevention of rug pulls. We analyzed the current taxonomy of rug pulls, combining the recent rug pull instances. Then, we developed a comprehensive taxonomy of rug pulls, which serves as an evaluation framework for our study.


\subsubsection{\textbf{RQ2}} \textit{What is the taxonomy coverage of the current Rug pull data set?} The progress in rug pull research and the creation of effective detection methods are hindered by the lack of comprehensive and dependable datasets. There is an unmet need for a thorough investigation into the current availability of data for training and testing these detection tools. It is crucial to assess the existing datasets, evaluate their scope, and determine how well they align with a detailed taxonomy of rug pulls.

\subsubsection{\textbf{RQ3}} \textit{What are the capabilities and limitations of current detection tools?} Detecting rug pulls is a demanding process that necessitates specific expert knowledge and considerable time investment. While there have been attempts to introduce automated tools for this purpose, a comprehensive understanding of their strengths and weaknesses remains lacking. Consequently, conducting an in-depth analysis of these tools is essential for their refinement, aiming to improve their precision, dependability, and applicability across a broader range of scenarios.

%% file: results_and_insights.tex
\section{Results and Insights}


\begin{table*}
\renewcommand{\arraystretch}{1.5} 
\caption{Taxonomy of Rug Pull Root Causes in Blockchain}
\label{tab:Taxonomy}
\begin{center}

\scalebox{0.7}{
\begin{tabular}{@{}llllll@{}}
\toprule
\textbf{Categories} &
  \textbf{Type} &
  \textbf{Root Cause} &
  \textbf{Scholar Paper~(SL)} &
  \textbf{Grey Literature~(GL)} &
  \textbf{Losses} \\ \midrule
\multirow{4}{*}{Simple Rug Rull} &
  \multirow{4}{*}{LP Drain} &
  \multirow{4}{*}{—} &
  \multirow{4}{*}{SL-\cite{An_analysis_of_crypto_scams_during_the_Covid-19_pandemic}, SL-\cite{Do_Not_Rug_on_Me}, SL-\cite{Rug_pull_malicious_token_detection_on_blockchain_using_supervised_learning_with_feature_engineering}} &
  \multirow{2}{*}{\begin{tabular}[c]{@{}l@{}}GL-\cite{BankrateRugPull2023}, GL-\cite{HackenRugPull2023}, GL-\cite{NerdWalletRugPull2023}, GL-\cite{ImmuneBytesRugPull2023}, \\ GL-\cite{BuiltInRugPull2023}, GL-\cite{CryptoVantageRugPulls2023}, GL\cite{ForkastNewsRugPull2023}, GL-\cite{NumenCyberRugPulls2023}, \\ GL-\cite{CoinInsiderRugPulls2023},  GL-\cite{HackerNoonSlowRugPulls2023}, GL-\cite{FrontalRugPull2023}, GL-\cite{ApeSpaceRugPull2023}, \\ GL-\cite{CoinscreedDeFiRugPulls2023}, GL-\cite{FinderCryptoRugPull2023}, GL-\cite{QuillAuditsRugPullDueDiligence2023}, GL-\cite{CertiKRugPullReport2023}\end{tabular}} &
  \multirow{2}{*}{\$5,223,217} \\  \rule{0pt}{38pt}
 &
   &
   &
   &
   &
   \\
Sell Rug Pull &
  Token Distribution &
  — &
  SL-\cite{TokenAuditor}, SL-\cite{Do_Not_Rug_on_Me}, SL-\cite{A_Deep_Dive_into_NFT_Rug_Pulls} &
  GL-\cite{ForkastNewsRugPull2023}, GL-\cite{QuillAuditsRugPullDueDiligence2023} &
  \$60,587,545 \\
SC Trap Doors &
  Token Generation &
  Mint &
  SL-\cite{ma2023pied}, SL-\cite{r3}, SL-\cite{Do_Not_Rug_on_Me} &
  GL-\cite{ISACASerialRugPull2022}, GL-\cite{CheckPointScammers2022}, GL-\cite{QuillAuditsScamTokenContracts2023} &
  \$19,465,791 \\
 &
   &
  Hidden Mint &
  SL-\cite{Scam_Alert}, SL-\cite{A_Deep_Dive_into_NFT_Rug_Pulls} &
  GL-\cite{soliduslabs_2023} &
   \\
 &
  Destroy Token &
  — &
  SL-\cite{ma2023pied} &
  GL-\cite{CheckPointScammers2022} &
   \\
 &
   &
  Burn* &
  — &
  GL-\cite{QuillAuditsScamTokenContracts2023} &
   \\
 &
  Transaction Limitation &
  Transfer Block &
  \begin{tabular}[c]{@{}l@{}}SL-\cite{ma2023pied}, SL-\cite{Scam_Alert}, SL-\cite{Do_Not_Rug_on_Me}, \\ SL-\cite{From_Programming_Bugs_to_Multimillion_Dollar_Scams}\end{tabular} &
  GL-\cite{ISACASerialRugPull2022}, GL-\cite{uniswap_2023} &
  \$127,500 \\
 &
   &
  Trading Cooldown &
  SL-\cite{ma2023pied} &
  GL-\cite{uniswap_2023} &
   \\
 &
   &
  Sale-Restrict &
\begin{tabular}[c]{@{}l@{}}SL-\cite{r3}, SL-\cite{An_analysis_of_crypto_scams_during_the_Covid-19_pandemic}, SL-\cite{From_Programming_Bugs_to_Multimillion_Dollar_Scams}, \\ SL-\cite{torres2019art}, SL-\cite{Why_Trick_Me}\end{tabular} &
  \begin{tabular}[c]{@{}l@{}}GL-\cite{soliduslabs_2023}, GL-\cite{CyberTalkDeFiRugPulls2021}, GL-\cite{BankrateRugPull2023}, GL-\cite{HackenRugPull2023}, \\ GL-\cite{ImmuneBytesRugPull2023}, GL-\cite{BuiltInRugPull2023}, GL-\cite{CryptoVantageRugPulls2023}, GL-\cite{ForkastNewsRugPull2023}, \\ GL-\cite{CheckPointScammers2022}, GL-\cite{NumenCyberRugPulls2023}, GL-\cite{CoinInsiderRugPulls2023}, GL-\cite{HackerNoonSlowRugPulls2023}, \\ GL-\cite{FrontalRugPull2023}, GL-\cite{ApeSpaceRugPull2023}, GL-\cite{CoinscreedDeFiRugPulls2023}, GL-\cite{FinderCryptoRugPull2023}, \\ GL-\cite{QuillAuditsRugPullDueDiligence2023}, GL-\cite{QuillAuditsScamTokenContracts2023}\end{tabular} &
   \\
 &
   &
  Freeze Account &
  SL-\cite{ma2023pied} &
  — &
   \\
 &
   &
  \begin{tabular}[c]{@{}l@{}}Transaction Number \\ or Amount Limitation\end{tabular} &
  SL-\cite{Why_Trick_Me} &
  GL-\cite{soliduslabs_2023}, GL-\cite{ISACASerialRugPull2022}, GL-\cite{uniswap_2023} &
   \\
 &
   &
  Whitelist &
  SL-\cite{TokenAuditor}, SL-\cite{Why_Trick_Me} &
  GL-\cite{soliduslabs_2023}, GL-\cite{uniswap_2023} &
   \\
 &
   &
  Blacklist &
  SL-\cite{TokenAuditor}, SL-\cite{Why_Trick_Me} &
  GL-\cite{soliduslabs_2023}, GL-\cite{ISACASerialRugPull2022}, GL-\cite{QuillAuditsScamTokenContracts2023}, GL-\cite{uniswap_2023} &
   \\
 &
  Fee &
  Fee Modification &
  SL-\cite{Scam_Alert}, SL-\cite{From_Programming_Bugs_to_Multimillion_Dollar_Scams}, SL-\cite{Why_Trick_Me} &
  GL-\cite{soliduslabs_2023}, GL-\cite{uniswap_2023} &
  \$60,000 \\
 &
   &
  High Sell Fee &
  SL-\cite{Why_Trick_Me} &
  GL-\cite{CheckPointScammers2022}, GL-\cite{QuillAuditsScamTokenContracts2023}, GL-\cite{uniswap_2023} &
   \\
 &
   &
  High Buy Fee &
  SL-\cite{Why_Trick_Me} &
  GL-\cite{CheckPointScammers2022}, GL-\cite{QuillAuditsScamTokenContracts2023} &
   \\
 &
   &
  Advance-Fee &
  SL-\cite{r3}, SL-\cite{A_Deep_Dive_into_NFT_Rug_Pulls} &
  — &
   \\
 &
   &
  Hidden Fee &
  SL-\cite{Why_Trick_Me} &
  — &
   \\
 &
  Funds Manipulation &
  Balance Modification &
  SL-\cite{Scam_Alert} &
  GL-\cite{soliduslabs_2023}, GL-\cite{uniswap_2023} &
  \$2,015,574 \\
 &
   &
  Fake Money Transfer &
  SL-\cite{The_greed_trap} &
  — &
   \\
 &
   &
  Arbitrarily Transfer &
  SL-\cite{ma2023pied}, SL-\cite{A_Deep_Dive_into_NFT_Rug_Pulls} &
  GL-\cite{soliduslabs_2023}, GL-\cite{Abubakar_Maruf_O.2022} &
   \\
 &
  Ownership Fraud &
  \begin{tabular}[c]{@{}l@{}}Fake Ownership \\ Renounce\end{tabular} &
  SL-\cite{Scam_Alert}, SL-\cite{The_greed_trap} &
  GL-\cite{soliduslabs_2023} &
   \\
 &
   &
  Ownership Transfer* &
  — &
  GL-\cite{defi_rekt} &
   \\
 &
   &
  Hidden Owner* &
  — &
  GL-\cite{ISACASerialRugPull2022}, GL-\cite{uniswap_2023} &
   \\
 &
  Unverified Contract* &
  — &
  — &
  GL-\cite{Abubakar_Maruf_O.2022} &
  \$1,913,898 \\
 &
  External Call* &
  — &
  — &
  GL-\cite{soliduslabs_2023}, GL-\cite{solidityscan2022circulatebusd}, GL-\cite{QuillAuditsScamTokenContracts2023} &
  \$7,779,182 \\
 &
  Proxy &
  — &
  SL-\cite{dos2022new} &
  GL-\cite{uniswap_2023} &
  \$39,686 \\
 &
  Vulnerability &
  — &
  SL-\cite{Do_Not_Rug_on_Me} &
  GL-\cite{CheckPointScammers2022}, GL-\cite{AnciliaUniswapAttack2023} &
  \$10,000 \\
LP Manipulation &
  Liquidity Pool Block &
  — &
  SL-\cite{Scam_Alert} &
  GL-\cite{soliduslabs_2023} &
   \\
 &
  Fake LP Lock* &
  — &
  — &
  GL-\cite{twitter2024certikalert} &
   \\
 &
  Wash-Trading &
  — &
  SL-\cite{r4}, SL-\cite{A_Deep_Dive_into_NFT_Rug_Pulls} &
  — &
   \\
 &
  Pump and Dump &
  — &
  SL-\cite{An_analysis_of_crypto_scams_during_the_Covid-19_pandemic}, SL-\cite{r4}, SL-\cite{Rug_pull_malicious_token_detection_on_blockchain_using_supervised_learning_with_feature_engineering} &
  \begin{tabular}[c]{@{}l@{}}GL-\cite{CyberTalkDeFiRugPulls2021}, GL-\cite{BankrateRugPull2023}, GL-\cite{HackenRugPull2023}, GL-\cite{NerdWalletRugPull2023}, \\ GL-\cite{ImmuneBytesRugPull2023}, GL-\cite{BuiltInRugPull2023}, GL-\cite{CryptoVantageRugPulls2023}, GL-\cite{NumenCyberRugPulls2023}, \\ GL-\cite{CoinInsiderRugPulls2023}, GL-\cite{HackerNoonSlowRugPulls2023}, GL-\cite{FrontalRugPull2023}, GL-\cite{ApeSpaceRugPull2023}, \\ GL-\cite{CoinscreedDeFiRugPulls2023}, GL-\cite{FinderCryptoRugPull2023}\end{tabular} &
   \\
 &
  Hedge &
  — &
  SL-\cite{r4} &
  — &
   \\
Counterfeit   Token &
  — &
  — &
  SL-\cite{torres2019art}, SL-\cite{A_Deep_Dive_into_NFT_Rug_Pulls}, SL-\cite{xu2023sok} &
  — &
   \\
Combination &
  — &
  — &
  SL-\cite{r4} &
  — &
  \$1,742,786 \\ \bottomrule
\end{tabular}}
\end{center}
\end{table*}

\subsection{RQ1: What is the taxonomy of root cause of rug pull?}

Our research has culminated in the development of the most comprehensive taxonomy of rug pulls to date. Drawing from an extensive review of 58 relevant literature sources, including 31 academic papers and 27 documents from grey literature, we have identified and categorized 34 distinct types of rug pulls into 6 high-level categories and 19 root cause categories as shown in \tabref{Taxonomy}. This taxonomy significantly expands upon the most widely cited study in the rug pull domain~\cite{r3}, introducing 15 new types and 31 new root cause categories, with 7 of these root cause categories being previously unacknowledged in academic literature. 

A notable feature of our taxonomy is the quantification of financial losses associated with each type of rug pull. By collecting data from case studies in grey literature and the Rekt~\cite{rekt-database}, we have determined that the total financial impact of rug pulls on users amounts to \textdollar171,713,065. Among the various categories, Token Distribution stands out as the most financially damaging, accounting for 35\% of the total losses. 
We introduce the taxonomy's first level depicting the various categories of rug pull operations:
\subsubsection{Simple rug pull} it occurs when a developer creates a new cryptocurrency token and pairs it with a high-value asset in a liquidity pool on a decentralized exchange to lure investors. Subsequently, the developer withdraws all the assets from said liquidity pool, rendering the token valueless and securing profit from the assets removed\cite{Do_Not_Rug_on_Me,An_analysis_of_crypto_scams_during_the_Covid-19_pandemic,Rug_pull_malicious_token_detection_on_blockchain_using_supervised_learning_with_feature_engineering}

\subsubsection{Sell rug pull} it typically entails the creation and manipulation of an extensive supply of a novel token, attracting investors through the appearance of secure trading pairs and eventually manipulating the market through the sale of vast quantities of the token, precipitating its value's collapse\cite{Do_Not_Rug_on_Me,TokenAuditor,A_Deep_Dive_into_NFT_Rug_Pulls}.

\subsubsection{Smart Contract(SC) Trap Doors} they are hidden or unintended functionalities that can be exploited to manipulate the contract's outcomes or misappropriate assets. These vulnerabilities can stem from deliberate malicious insertions by the creator, coding errors, complex interactions with other contracts or data sources, or through features intended for upgrading the contract\cite{Do_Not_Rug_on_Me}\cite{From_Programming_Bugs_to_Multimillion_Dollar_Scams}\cite{TokenAuditor}.
This category is composed of the following various subtypes:

\paragraph{Token Generation} it centers on the risks inherent to the minting of new tokens, a foundational process in numerous DeFi projects and ecosystems. If not securely architectured and implemented, this process can expose the system to vulnerabilities and threats. Risks fall into two main categories: \textit{mint}, wherein issuers generate new tokens under specific conditions that may be exploited to covertly inflate balances and undermine trust\cite{ma2023pied,r3,Do_Not_Rug_on_Me}, and \textit{hidden mint}, which entails the surreptitious creation of tokens beyond the publicly stated maximum cap, thereby stealthily manipulating the supply sans community awareness\cite{Scam_Alert,A_Deep_Dive_into_NFT_Rug_Pulls}. \paragraph{Destroy Token} it includes backdoors enabling the obliteration of tokens from any account at will\cite{ma2023pied} and malicious \textit{burn} functions that let owners erase tokens from users' wallets, artificially inflating value for exploitation\cite{quillaudit_2022}.
\paragraph{Transaction limitation} it includes \textit{transfer block} (halting all transfers)\cite{ma2023pied}\cite{Scam_Alert}\cite{A_Deep_Dive_into_NFT_Rug_Pulls}\cite{From_Programming_Bugs_to_Multimillion_Dollar_Scams}, \textit{trading cooldown} (delaying sales post-purchase)\cite{ma2023pied}, \textit{sale-restrict} (allowing only creators to sell)\cite{r3}\cite{An_analysis_of_crypto_scams_during_the_Covid-19_pandemic}\cite{From_Programming_Bugs_to_Multimillion_Dollar_Scams}\cite{torres2019art}\cite{Why_Trick_Me}, \textit{freeze account} (blocking specific wallets)\cite{ma2023pied}, \textit{transaction number or amount limitation} (restricting transfer amounts or numbers)\cite{Why_Trick_Me}, and \textit{whitelist/blacklist} mechanisms (selectively permitting or denying transactions)\cite{TokenAuditor}\cite{Why_Trick_Me}. \paragraph{Fee} it encompasses \textit{fee modification} (enabling sudden, exorbitant fee increases on transactions)\cite{uniswap_2023}\cite{soliduslabs_2023}, \textit{high sell/buy fee} (skewing token exchange rates unfavorably)\cite{Why_Trick_Me}, \textit{advance-fee} (charging upfront fees on secondary market trades, potentially preceding rug pulls)\cite{r3}\cite{A_Deep_Dive_into_NFT_Rug_Pulls}, and \textit{hidden fee} (untransparent charges reducing trade value)\cite{Why_Trick_Me}\cite{A_Deep_Dive_into_NFT_Rug_Pulls}. \paragraph{Funds manipulation} it includes \textit{balance modification} (enabling unauthorized changes to users' token balances)\cite{uniswap_2023}\cite{soliduslabs_2023}, \textit{fake money transfer}\cite{A_Deep_Dive_into_NFT_Rug_Pulls}, and \textit{arbitrary transfer} (allowing unauthorized redirection of tokens from any address)\cite{ma2023pied}\cite{A_Deep_Dive_into_NFT_Rug_Pulls}. \paragraph{Ownership fraud} it includes \textit{fake ownership renounce} (creators pretending to relinquish control while retaining special access)\cite{Scam_Alert}\cite{The_greed_trap}, \textit{ownership transfer} (shifting control of the contract to another party)\cite{defi_rekt}, and \textit{hidden owner} (developers maintaining the ability to manipulate contracts post-ownership renouncement)\cite{phan2022serialrugpull}\cite{uniswap_2023}. \paragraph{Unverified contract} it means the contract's source code has not been made public or verified on blockchain explorers, preventing users from reviewing its functions and potential vulnerabilities\cite{Abubakar_Maruf_O.2022}.
\paragraph{External call} it involves a token's transfer functionality being managed by a separate, private contract whose source code is hidden\cite{soliduslabs_2023}\cite{quillaudits2022backdoors}\cite{solidityscan2022circulatebusd}. 
\paragraph{Proxy contract}  it allows the owner to alter token functions, potentially impacting its price\cite{uniswap_2023}. \paragraph{Vulnerability} it exploits flaws in smart contract design, particularly in how tokens interact with decentralized exchanges like Uniswap\cite{Do_Not_Rug_on_Me}.

\subsubsection{LP manipulation} it encompasses \textit{Liquidity Pool Blocks} (restricting transfers from pools)\cite{Scam_Alert}, \textit{Fake LP Lock} (owners selling supposedly locked tokens)\cite{twitter2024certikalert}, \textit{Wash-trading} (faking trading volume through circular transactions)\cite{r4}\cite{A_Deep_Dive_into_NFT_Rug_Pulls}, \textit{Pump and Dump} (inflating prices for profit before selling off)\cite{An_analysis_of_crypto_scams_during_the_Covid-19_pandemic,r4}, and \textit{Hedging} (selling reserved tokens at peak prices)\cite{Rug_pull_malicious_token_detection_on_blockchain_using_supervised_learning_with_feature_engineering}.

\subsubsection{Counterfeit Token} it involves creating tokens that closely mimic reputable projects, aiming to mislead and exploit investors by capitalizing on the established name's trust and recognition\cite{From_Programming_Bugs_to_Multimillion_Dollar_Scams}\cite{A_Deep_Dive_into_NFT_Rug_Pulls}\cite{xu2023sok}.

\subsubsection{Combination} the aforementioned manipulation techniques often deployed together by scammers, strategically integrating them to maximize their deceptive impact and achieve their fraudulent objectives\cite{r4}.


\answer{1}{ We developed a taxonomy that classifies 34 types of rug pulls into six high-level and 19 root causes categories while also quantifying the financial impact associated with each type.}

\begin{table*}
\centering
\caption{Tools and Datasets Root Cause Coverage of Rug Pull}
\centering
\scalebox{0.5}{
\large
\begin{tabular}{@{}lllllllllllllll|llll@{}}
\toprule
Root Causes &
  T-01 &
  T-02 &
  T-03 &
  T-04 &
  T-05 &
  T-06 &
  T-07 &
  T-08 &
  T-09 &
  T-10 &
  T-11 &
  T-12 &
  T-13 &
  T-14 &
  D-01 &
  D-02 &
  D-03 &
  D-04 \\ \midrule
LP Drain             &   &   & $\blacksquare$ &   &   &   &   &   &   &   &   &   &   &   &        &  &       & $\blacksquare$(7)   \\
Token Distribution   & $\blacksquare$ &   & $\blacksquare$ & $\blacksquare$ &   & $\blacksquare$ & $\blacksquare$ & $\blacksquare$ & $\blacksquare$ & $\blacksquare$ & $\blacksquare$ & $\blacksquare$ & $\blacksquare$ &   &        &  &       & $\blacksquare$(13)  \\
Mint                 & $\blacksquare$ & $\blacksquare$ & $\blacksquare$ & $\blacksquare$ & $\blacksquare$ &   & $\blacksquare$ & $\blacksquare$ & $\blacksquare$ & $\blacksquare$ &   &   & $\blacksquare$ & $\blacksquare$ &        &  & $\blacksquare$(34) & $\blacksquare$(65)  \\
Hidden Mint          &   &   &   & $\blacksquare$ &   &   &   &   &   &   &   &   &   &   &        &  &       & $\blacksquare$(1)   \\
Destroy Token        &   &   &   &   &   &   &   &   &   &   &   &   &   &   &        &  &       &        \\
Burn                 & $\blacksquare$ & $\blacksquare$ &   & $\blacksquare$ & $\blacksquare$ &   &   & $\blacksquare$ &   &   &   &   &   &   &        &  & $\blacksquare$(29) & $\blacksquare$(30)  \\
Transfer Block       & $\blacksquare$ & $\blacksquare$ & $\blacksquare$ & $\blacksquare$ & $\blacksquare$ & $\blacksquare$ & $\blacksquare$ &   &   & $\blacksquare$ & $\blacksquare$ & $\blacksquare$ & $\blacksquare$ & $\blacksquare$ & $\blacksquare$(2)   &  & $\blacksquare$(29) & $\blacksquare$(31)  \\
Trading Cooldown     & $\blacksquare$ & $\blacksquare$ & $\blacksquare$ & $\blacksquare$ & $\blacksquare$ &   &   & $\blacksquare$ & $\blacksquare$ & $\blacksquare$ &   &   & $\blacksquare$ & $\blacksquare$ &        &  &       & $\blacksquare$(1)   \\
Sale-Restrict &
  $\blacksquare$ &
  $\blacksquare$ &
  $\blacksquare$ &
   &
   &
  $\blacksquare$ &
  $\blacksquare$ &
   &
  $\blacksquare$ &
  $\blacksquare$ &
  $\blacksquare$ &
  $\blacksquare$ &
  $\blacksquare$ &
  $\blacksquare$ &
  $\blacksquare$(1701) &
  $\blacksquare$(282) &
   &
  $\blacksquare$(1985) \\
Freeze Account       &   &   &   &   &   &   &   &   &   &   &   &   &   &   &        &  & $\blacksquare$(95) & $\blacksquare$(95)  \\
\begin{tabular}[c]{@{}l@{}}Transaction Number \\ or Amount Limitation\end{tabular} &
  $\blacksquare$ &
  $\blacksquare$ &
  $\blacksquare$ &
  $\blacksquare$ &
  $\blacksquare$ &
  $\blacksquare$ &
   &
   &
  $\blacksquare$ &
  $\blacksquare$ &
  $\blacksquare$ &
  $\blacksquare$ &
  $\blacksquare$ &
  $\blacksquare$ &
   &
   &
   &
   \\
Whitelist            & $\blacksquare$ & $\blacksquare$ & $\blacksquare$ & $\blacksquare$ & $\blacksquare$ &   &   & $\blacksquare$ &   & $\blacksquare$ &   &   & $\blacksquare$ & $\blacksquare$ &        &  &       &        \\
Blacklist            & $\blacksquare$ & $\blacksquare$ & $\blacksquare$ & $\blacksquare$ & $\blacksquare$ & $\blacksquare$ & $\blacksquare$ & $\blacksquare$ & $\blacksquare$ & $\blacksquare$ & $\blacksquare$ &   & $\blacksquare$ & $\blacksquare$ &        &  &       & $\blacksquare$(1)   \\
Fee Modification     & $\blacksquare$ & $\blacksquare$ & $\blacksquare$ & $\blacksquare$ & $\blacksquare$ & $\blacksquare$ & $\blacksquare$ & $\blacksquare$ & $\blacksquare$ & $\blacksquare$ & $\blacksquare$ &   & $\blacksquare$ &   & $\blacksquare$(154) &  &       & $\blacksquare$(155) \\
High Sell Fee        & $\blacksquare$ & $\blacksquare$ & $\blacksquare$ & $\blacksquare$ &   & $\blacksquare$ & $\blacksquare$ &   & $\blacksquare$ & $\blacksquare$ & $\blacksquare$ & $\blacksquare$ & $\blacksquare$ &   &        &  &       &        \\
High Buy Fee         & $\blacksquare$ & $\blacksquare$ & $\blacksquare$ & $\blacksquare$ &   & $\blacksquare$ & $\blacksquare$ &   & $\blacksquare$ & $\blacksquare$ & $\blacksquare$ & $\blacksquare$ & $\blacksquare$ &   &        &  &       &        \\
Advance-Fee          &   &   &   & $\blacksquare$ &   &   &   &   &   &   &   &   &   &   &        &  &       &        \\
Hidden Fee           &   &   &   &   &   &   &   &   &   &   &   &   &   &   &        &  &       &        \\
Balance Modification &   & $\blacksquare$ & $\blacksquare$ & $\blacksquare$ &   &   &   & $\blacksquare$ &   & $\blacksquare$ &   &   &   & $\blacksquare$ &        &  &       & $\blacksquare$(7)   \\
Fake Money Transfer  &   &   &   &   &   &   &   &   &   &   &   &   &   &   &        &  &       &        \\
Arbitrarily Transfer & $\blacksquare$ &   & $\blacksquare$ &   &   &   &   &   &   &   &   & $\blacksquare$ &   &   &        &  & $\blacksquare$(2)  & $\blacksquare$(6)   \\
\begin{tabular}[c]{@{}l@{}}Fake Ownership \\ Renounce\end{tabular} &
  $\blacksquare$ &
   &
  $\blacksquare$ &
   &
   &
  $\blacksquare$ &
   &
  $\blacksquare$ &
   &
  $\blacksquare$ &
   &
   &
  $\blacksquare$ &
  $\blacksquare$ &
   &
   &
   &
   \\
Ownership Transfer   &   &   &   &   &   &   &   &   &   &   &   &   &   &   &        &  &       &        \\
Hidden Owner         &   & $\blacksquare$ & $\blacksquare$ & $\blacksquare$ &   &   &   &   &   & $\blacksquare$ &   &   &   & $\blacksquare$ &        &  &       &        \\
Unverified Contract  & $\blacksquare$ & $\blacksquare$ & $\blacksquare$ & $\blacksquare$ & $\blacksquare$ & $\blacksquare$ &   & $\blacksquare$ & $\blacksquare$ & $\blacksquare$ &   & $\blacksquare$ & $\blacksquare$ & $\blacksquare$ &        &  &       & $\blacksquare$(1)   \\
External Call        & $\blacksquare$ &   & $\blacksquare$ & $\blacksquare$ &   &   &   &   & $\blacksquare$ & $\blacksquare$ &   &   &   & $\blacksquare$ &        &  &       & $\blacksquare$(7)   \\
Proxy                & $\blacksquare$ & $\blacksquare$ & $\blacksquare$ & $\blacksquare$ & $\blacksquare$ & $\blacksquare$ &   & $\blacksquare$ & $\blacksquare$ & $\blacksquare$ &   & $\blacksquare$ & $\blacksquare$ & $\blacksquare$ &        &  &       & $\blacksquare$(1)   \\
Vulnerability        &   & $\blacksquare$ & $\blacksquare$ &   &   &   &   &   &   &   &   &   &   &   &        &  &       & $\blacksquare$(2)   \\
Liquidity Pool Block &   &   &   &   &   &   &   &   &   &   &   &   &   &   &        &  &       &        \\
Fake LP Lock         &   &   &   &   &   &   &   &   &   &   &   &   &   &   &        &  &       & $\blacksquare$(1)   \\
Wash-Trading         &   &   &   &   &   &   &   &   &   &   &   &   &   &   &        &  &       &        \\
Pump and Dump        &   &   & $\blacksquare$ &   &   &   &   &   &   &   &   &   &   &   &        &  &       &        \\
Hedge                &   &   &   &   &   &   &   &   &   &   &   &   &   &   &        &  &       &        \\
Counterfeit Token    & $\blacksquare$ & $\blacksquare$ &   &   &   &   &   &   &   &   &   &   &   &   &        &  &       &        \\
Combination          & $\blacksquare$ & $\blacksquare$ & $\blacksquare$ & $\blacksquare$ & $\blacksquare$ & $\blacksquare$ & $\blacksquare$ & $\blacksquare$ &  $\blacksquare$ & $\blacksquare$ & $\blacksquare$ & $\blacksquare$ & $\blacksquare$ & $\blacksquare$ &        &  &       & $\blacksquare$(15)  \\ \bottomrule
\end{tabular}}
\label{tab:Coverage}
\end{table*}

\subsection{RQ2: What is the taxonomy coverage of the current Rug pull data set?}

Through a systematic evaluation of 31 academic papers, we identified three datasets that explicitly include annotated rug pull instances, primarily aimed at facilitating the development of rug pull detection tools. These datasets, namely Trapdoor-D \cite{From_Programming_Bugs_to_Multimillion_Dollar_Scams}, Honeypot-D \cite{torres2019art}, and Backdoor-D \cite{ma2023pied}, collectively encompass 2,508 instances of rug pulls across various platforms including Ethereum and Binance.

However, our analysis revealed that these existing datasets have limited coverage of the diverse landscape of rug pulls. As illustrated in \tabref{Coverage}, the three datasets combined only cover 7 out of the 34 identified rug pull root causes, resulting in a mere 20\% overall coverage. Furthermore, there is significant variance in the number of instances per rug pull type, with ``Sale-Restrict" being the most prevalent (1,701 instances) and ``Arbitrarily Transfer" being the least represented (2 instances). The insufficient coverage of rug pull operations may arise from their multifaceted nature in practical settings, coupled with the challenges involved in detecting specific types of pulls, notably those that manipulate liquidity pools.

To address these limitations, we augmented the dataset by integrating the existing datasets with 90 real-world rug pull instances sourced from grey literature.
The resulting reconstructed dataset, comprising 2,360 instances across 13 root cause categories, offers a more diverse and comprehensive representation of the rug pull landscape. 
Our efforts have increased the coverage of rug pull root causes from 7 to 19 types, boosting the overall coverage rate from 20\% to 56\%. The lack of 16 rug pull types (46\%) in the data set may be due to the lack of a comprehensive and unified rug pull type definition in the industry, as well as the diversity and complexity of rug pull methods in the real world. This enhanced dataset provides a solid foundation for further research and analysis in the field of rug pull detection and mitigation.

\answer{2}{The three available datasets contain a total of 2,508 instances, covering only 7 different rug pull root causes, resulting in a 20\% coverage rate (7 out of 34 types). Our reconstructed dataset contains 2,360 instances after removing duplicated addresses, covering 19 different rug pull root causes, with an overall coverage rate of 56\% (19 out of 34 types).}

\begin{table}
\caption{Dataset of Rug Pull in Blockchain}
\centering
\scalebox{0.75}{
\begin{tabular}{@{}lllllll@{}}
\toprule
Dataset &
  \begin{tabular}[c]{@{}l@{}}Publish\\ Year\end{tabular} &
  Chain &
  \# Instances &
  \begin{tabular}[c]{@{}l@{}}\# Type\\ coverage\end{tabular} &
  \begin{tabular}[c]{@{}l@{}}\# Root Causes\\ coverage\end{tabular} \\ \midrule
Trapdoor-D\cite{From_Programming_Bugs_to_Multimillion_Dollar_Scams}  & 2023 & Ethereum                                                    & 1859 & 2/19  & 3/34  \\
Honeypot-D\cite{torres2019art}  & 2019 & Ethereum                                                    & 460  & 1/19  & 1/34  \\
Backdoor-D\cite{ma2023pied}  & 2022 & Ethereum                                                    & 189  & 4/19  & 5/34  \\
Our-Dataset\cite{OurRugPullDataset} & 2024 & \begin{tabular}[c]{@{}l@{}}Ethereum,\\ Binance\end{tabular} & 2360 & 13/19 & 19/34 \\ \bottomrule
\end{tabular}}
\label{table:Dataset of Rug Pull in Blockchain}
\end{table}

\subsection{RQ3: What are the capabilities and limitations of current detection tools?} 
We identified 14 tools that specifically aim to identify rug pulls and pinpoint their root causes from the selected 58 pieces of literature as shown in \tabref{Final tools list}. 

\begin{table}
\centering
\caption{Final Tools List}
\label{tab:Final tools list}
\scalebox{0.8}{
\begin{tabular}{@{}lll@{}}
\toprule
Tools & Name          & Urls                                              \\ \midrule
T-01  & GoPlus        & https://gopluslabs.io/                            \\
T-02  & Aegisweb3     & https://www.aegisweb3.com/                        \\
T-03  & De.Fi         & https://de.fi/                                    \\
T-04  & Quick Intel   & https://app.quickintel.io/scanner                 \\
T-05  & Cyberscan     & https://www.cyberscope.io/cyberscan               \\
T-06  & Token Sniffer & https://tokensniffer.com/                         \\
T-07  & Staysafu      & https://app.staysafu.org/                         \\
T-08  & Solidityscan  & https://solidityscan.com/                         \\
T-09  & Contractwolf  & https://contractwolf.io/rugscanner                \\
T-10  & Blocksafu     & https://blocksafu.com/token-scanner               \\
T-11  & Dexanalyzer   & https://www.dexanalyzer.io/                       \\
T-12  & Honeypot.is   & https://honeypot.is/                              \\
T-13  & Hapilabs      & https://terminal.hapilabs.one/guest-address-check \\
T-14  & Quillcheck    & https://quillcheck.quillaudits.com/               \\ \bottomrule
\end{tabular}}
\end{table}

We employ the developed taxonomy in \tabref{Taxonomy} to evaluate these tools.
The results, as presented in \tabref{Coverage}, reveal that collectively, these tools can detect 25 out of the 34 identified rug pull root causes, achieving a coverage rate of 73.5\%. However, there are notable variations in the performance of individual tools, with De.Fi (T-03) exhibits the highest coverage rate of 61.76\% (21 out of 34 root causes), and Staysafu (T-07) and Dexanalyzer (T-11) have the lowest coverage rate of 23.53\% (8 out of 34 root causes).

Despite the collective coverage achieved by these tools, our analysis highlights significant gaps in their detection capabilities. Notably, 9 rug pull root causes (accounting for 36\% of the total) remain undetected by any of the tools. This finding underscores the existence of undiscovered root causes and potential areas for further research and tool development. 

To gauge the real-world effectiveness of these tools, we conducted an empirical study on 30 verified instances from our comprehensive dataset, encompassing both simple~(single root cause) and complex~(combined root causes) rug pull scenarios (Tables \ref{Real World Incidents Results} and \ref{Real World Combination Incidents Results}). The results reveal significant limitations in the tools' performance under practical conditions. 
For single rug pull root cause type, the highest detection hit rate was 100\% (Dexanalyzer, T-11), while two tools (Hapilabs, T-03 and Staysafu, T-07) had a hit rate of 0\%. In the case of compound rug pull root cause types, GoPlus (T-01) performed the best with an average hit rate of 50\%, followed by Blocksafu (T-10) at 45\%, while three tools (Token Sniffer, T-06; Hapilabs, T-03; and Staysafu, T-07) had an average hit rate of 0\%. 
Current rug pull detection tools exhibit significant limitations, reflecting a lack of comprehensiveness and robustness in addressing the diverse and evolving landscape of rug pulls in the cryptocurrency ecosystem.

\begin{table*}
\centering
\caption{Real World Incidents Results}
\label{Real World Incidents Results}
\begin{threeparttable}
\begin{tabular}{@{}lllllllllllllll@{}}
\toprule
Root Causes          & T-01 & T-02 & T-03 & T-04 & T-05 & T-06 & T-07 & T-08 & T-09 & T-10 & T-11 & T-12 & T-13 & T-14 \\ \midrule
Sale-Restrict        & 0    & 0    & 0    & *    & *    & 0    & NaN  & *    & 0    & 0    & *    & 0    & *    & 0    \\
Destroy Token        & *    & *    & *    & *    & *    & *    & *    & *    & *    & *    & *    & *    & NaN  & *    \\
Trading Cooldown     & 1    & 1    & 0    & 1    & 1    & *    & *    & 1    & 1    & 1    & *    & *    & *    & 1    \\
Token Distribution   & 1    & *    & 1    & 1    & 0    & 0    & NaN  & 1    & 1    & 1    & 1    & *    & NaN  & 1    \\
Mint                 & 1    & 1    & 1    & 1    & 1    & 0    & NaN  & 1    & 1    & 1    & *    & *    & *    & 0    \\
External Call        & 1    & *    & 1    & 1    & *    & *    & *    & *    & 1    & 1    & *    & *    & *    & 1    \\
LP Drain             & *    & *    & 1    & *    & *    & *    & *    & *    & *    & *    & *    & *    & NaN  & *    \\
Fake LP Lock         & *    & *    & *    & *    & *    & *    & *    & *    & *    & *    & *    & *    & NaN  & *    \\
Balance Modification & 0    & 0    & 0    & 1    & *    & *    & *    & 0    & *    & 0    & *    & *    & *    & *    \\
Arbitrarily Transfer & NaN  & NaN  & 0    & *    & *    & *    & *    & *    & NaN  & NaN  & *    & NaN  & NaN  & *    \\
Blacklist            & 1    & 1    & 1    & 1    & 1    & 1    & NaN  & 1    & 1    & 1    & 1    & *    & *    & 1    \\
Fee Modification     & 1    & 1    & 0    & 1    & 1    & 1    & NaN  & 0    & *    & 1    & 1    & *    & NaN  & 1    \\
Proxy                & 1    & 1    & 0    & 0    & 0    & NaN  & *    & 0    & 1    & 1    & 1    & 1    & *    & 0    \\
Vulnerability        & *    & 1    & NaN  & *    & *    & *    & *    & \$   & *    & \$   & *    & *    & NaN  & *    \\
Hidden Mint          & 1    & 0    & 0    & 0    & 0    & 0    & NaN  & 0    & 1    & 1    & *    & *    & *    & 0    \\ \bottomrule
\end{tabular}
\begin{tablenotes}
    \footnotesize
    \item "NaN" signifies that the tool is incapable of displaying the output. \\"*" indicates that the tool lacks support for the specified type of root cause. \\"\$" denotes a requirement for proprietary tools, which necessitate payment, to support the diagnosis of this root cause category.
  \end{tablenotes}
\end{threeparttable}
\end{table*}

\begin{table*}
\centering
\caption{Real World Combination Incidents Results}
\label{Real World Combination Incidents Results}
\begin{threeparttable}

\begin{tabular}{@{}lllllllllllllll@{}}
\toprule
Combination Root Causes &
  T-01 &
  T-02 &
  T-03 &
  T-04 &
  T-05 &
  T-06 &
  T-07 &
  T-08 &
  T-09 &
  T-10 &
  T-11 &
  T-12 &
  T-13 &
  T-14 \\ \midrule
Balance Modification, Burn &
  1/2 &
  0/2 &
  1/2 &
  1/2 &
  0/2 &
  0/2 &
  NaN &
  0/2 &
  0/2 &
  1/2 &
  0/2 &
  0/2 &
  NaN &
  0/2 \\
Mint, Sale-Restrict &
  1/2 &
  0/2 &
  2/2 &
  1/2 &
  0/2 &
  0/2 &
  NaN &
  1/2 &
  1/2 &
  1/2 &
  1/2 &
  1/2 &
  NaN &
  1/2 \\
\begin{tabular}[c]{@{}l@{}}Arbitrarily Transfer, Mint,   \\ Burn, Fee Modification\end{tabular} &
  2/4 &
  1/4 &
  1/4 &
  1/4 &
  1/4 &
  0/5 &
  NaN &
  2/4 &
  1/4 &
  2/4 &
  NaN &
  0/4 &
  NaN &
  1/4 \\
Hidden Mint, Hidden Owner &
  0/2 &
  0/2 &
  0/2 &
  0/2 &
  0/2 &
  0/2 &
  NaN &
  0/2 &
  0/2 &
  0/2 &
  0/2 &
  0/2 &
  NaN &
  0/2 \\
\begin{tabular}[c]{@{}l@{}}Balance Modification,\\ Sale-Restrict\end{tabular} &
  2/2 &
  0/2 &
  1/2 &
  0/2 &
  0/2 &
  0/2 &
  NaN &
  0/2 &
  1/2 &
  2/2 &
  0/2 &
  0/2 &
  NaN &
  0/2 \\
Hidden Mint, Blacklist &
  2/2 &
  0/2 &
  0/2 &
  0/2 &
  0/2 &
  0/2 &
  NaN &
  0/2 &
  2/2 &
  2/2 &
  0/2 &
  0/2 &
  NaN &
  0/2 \\
Hidden Mint, Blacklist &
  2/2 &
  0/2 &
  0/2 &
  0/2 &
  0/2 &
  0/2 &
  NaN &
  0/2 &
  2/2 &
  2/2 &
  0/2 &
  0/2 &
  NaN &
  0/2 \\
Mint, Balance Modification &
  1/2 &
  0/2 &
  0/2 &
  0/2 &
  0/2 &
  0/2 &
  NaN &
  0/2 &
  1/2 &
  1/2 &
  0/2 &
  0/2 &
  NaN &
  1/2 \\
\begin{tabular}[c]{@{}l@{}}Arbitrarily Transfer, \\ Whitelist\end{tabular} &
  1/2 &
  1/2 &
  1/2 &
  1/2 &
  1/2 &
  0/2 &
  NaN &
  1/2 &
  0/2 &
  0/2 &
  0/2 &
  0/2 &
  NaN &
  1/2 \\
Burn, Sale-Restrict &
  0/2 &
  1/2 &
  1/2 &
  1/2 &
  1/2 &
  0/2 &
  NaN &
  0/2 &
  0/2 &
  0/2 &
  0/2 &
  0/2 &
  NaN &
  1/2 \\
Fee Modification, Burn &
  0/2 &
  0/2 &
  NaN &
  1/2 &
  1/2 &
  0/2 &
  NaN &
  0/2 &
  0/2 &
  0/2 &
  1/2 &
  0/2 &
  NaN &
  1/2 \\
Mint, Balance Modification &
  2/2 &
  1/2 &
  NaN &
  0/2 &
  0/2 &
  0/2 &
  NaN &
  0/2 &
  1/2 &
  2/2 &
  0/2 &
  0/2 &
  NaN &
  1/2 \\
Sale-Restrict, Wash Trading &
  0/2 &
  1/2 &
  0/2 &
  0/2 &
  0/2 &
  Fail &
  NaN &
  0/2 &
  1/2 &
  0/2 &
  1/2 &
  1/2 &
  NaN &
  0/2 \\
Mint, Ownership Transfer &
  1/2 &
  0/2 &
  0/2 &
  0/2 &
  0/2 &
  0/2 &
  NaN &
  0/2 &
  1/2 &
  1/2 &
  0/2 &
  0/2 &
  NaN &
  0/2 \\
\begin{tabular}[c]{@{}l@{}}Token Distribution, \\ Advance-Fee\end{tabular} &
  1/2 &
  2/2 &
  2/2 &
  1/2 &
  0/2 &
  0/2 &
  NaN &
  1/2 &
  1/2 &
  1/2 &
  1/2 &
  0/2 &
  NaN &
  1/2 \\ \bottomrule
\end{tabular}

\begin{tablenotes}
    \footnotesize
    \item "NaN" signifies that the tool is incapable of displaying the output. 
  \end{tablenotes}
\end{threeparttable}
 \vspace*{-2em}
\end{table*}

\answer{3}{The 14 selected tools collectively identified 25 out of 34 rug pull root causes delineated in our taxonomy, achieving a coverage rate of 73.5\%. Nevertheless, 9 rug pull root causes, representing 26.5\% of the total, remained undetected by all the tools examined. In real-world scenarios, for singular rug pull root cause types, the maximum detection rate achieved by one tool was 100\%, while two tools exhibited a 0\% success rate. For the combination rug pull root cause type, no tool surpassed an average detection rate of 50\%, and three tools consistently failed to detect any, averaging a 0\% hit rate.}

%% file: discussion.tex
\section{Threads to Validity}
\subsection{\textit{Internal Threats}}There exists the potential for bias within the manual classification and annotation process. The development of the comprehensive taxonomy and the categorization of rug pull types from various sources relied on the judgment and expertise of the researchers involved. While efforts were made to ensure consistency and objectivity through collaborative decision-making and independent evaluations, there remains a risk of subjectivity influencing the results. 
\subsection{\textit{External Threats}}Although this study sought comprehensiveness by examining numerous academic and grey literature sources, it is conceivable that certain relevant rug pull methods or cases may have eluded capture. The dynamic blockchain and cryptocurrency landscape suggests that novel rug pull strategies could manifest, which may challenge the current taxonomy's pertinence to future conditions. Additionally, the datasets employed, while reconstructed and augmented, might not fully encapsulate the spectrum of rug pull cases across various blockchain infrastructures. To mitigate these external risks, it is imperative to perpetuate research endeavors that periodically refresh the taxonomy and datasets in response to emerging rug pull modalities and occurrences. Engaging with industry authorities and the broader blockchain fellowship can enhance the taxonomy's exhaustiveness and applicability.

%% file: related_work.tex
\section{Related Work}

Sharma et al.\cite{sharma2023understanding} delved into an expansive dataset—comprising 758 instances of rug pulls—across NFT marketplaces, analyzing structural and behavioral patterns to unravel the motivations behind such frauds. Kaur et al.\cite{kaur2023risk} employed a fuzzy analytical hierarchy process, uncovering that technical risks top the hierarchy within DeFi, followed by legal, regulatory, and financial risks. Cernera et al.\cite{r4} assessed the sustainability of tokens on both Ethereum and Binance Smart Chain, with a particular focus on the enduring effects of fraudulent schemes, such as rug pulls, and the insidious role of sniper bots. Aliyev et al.\cite{Scam_Alert} scrutinized ERC-20 tokens on decentralized exchanges, developing methods to identify and forecast rug pulls in liquidity pools.

On the basis of these empirical insights, scholars have advanced methods for the detection and curtailment of rug pull scams. Mazorra et al.\cite{Do_Not_Rug_on_Me} as well as Xia et al.\cite{r3} harnessed machine learning to pinpoint and classify fraudulent tokens on Uniswap, disclosing a staggering figure of over 10,000 scam tokens and emphasizing the widespread nature of rug pulls. Huang et al.\cite{A_Deep_Dive_into_NFT_Rug_Pulls} conducted the foremost comprehensive study on NFT rug pulls, devising a predictive model that integrates contract and transaction data with social media activity to proactively signal potential frauds. Nguyen et al.\cite{Rug_pull_malicious_token_detection_on_blockchain_using_supervised_learning_with_feature_engineering} crafted a supervised learning approach, supplemented with sophisticated feature engineering, to discern rug pull schemes in blockchain tokens. Cao et al.\cite{TokenAuditor} and Ma et al.\cite{ma2023pied} formulated innovative frameworks that employ fuzzing and hybrid analysis to detect flaws and exploitation risks in smart contracts, signaling the continuous pursuit of effective strategies for the detection and prevention of rug pull scams.

%% file: conclusion.tex
\section{Conclusion and Discussion}

We have conducted a comprehensive and systematic study on rug pulls from both academic and industrial perspectives. We established a systematic analytical research framework~(RPAF) and summarized an exhaustive rug pull taxonomy drawn from selected literature and books. It covers 34 types of rug pulls, their causes, and the resulting losses. Based on this taxonomy, we assessed the 3 rug pull datasets used in existing research and found significant discrepancies with actual conditions, prompting us to expand the datasets. The augmented dataset can benefit the research community. To further explore the current state of rug pull prevention, we applied our taxonomy to analyze and evaluate the existing 14 detection tools, identifying a critical need for improvement in their detection capabilities. This study bridges the gap between current rug pull research and practical application. We systematically summarized the rug pull taxonomy and analyzed its causes and the losses incurred. 

Our research indicates that the patterns of rug pulls are becoming increasingly complex. Scammers are shifting from initially simple methods to more deceptive, hybrid approaches. This transition makes detecting rug pulls more challenging. However, all rug pulls share a common goal: to drain the system's funds and exit with a profit. We should identify the objective to detect and uncover all potential methods that could instantaneously deplete the system's liquidity. Any means capable of abruptly draining the financial liquidity in the system is unacceptable in a healthy and positive system and contradicts the decentralized governance principles~\cite{ma2023comprehensive} of DeFi projects.